Positive Psychology and Romantic Scientism:

Reply to Comments on Brown, Sokal, & Friedman (2013)

Nicholas J. L. Brown
*New School of Psychotherapy and Counselling*

Alan D. Sokal
*New York University and University College London*

Harris L. Friedman
*Saybrook University and University of Florida*

September 17, 2014

**Running Head:** Positive Psychology and Romantic Scientism

**Author note**

Nicholas J. L. Brown, New School of Psychotherapy and Counselling, London, UK; Alan D. Sokal, Department of Physics, New York University and Department of Mathematics, University College London; Harris L. Friedman, School of Psychology and Interdisciplinary Inquiry, Saybrook University and Department of Psychology, University of Florida.

Correspondence concerning this article should be addressed to Alan D. Sokal, Department of Physics, New York University, 4 Washington Place, New York, NY 10003. E-mail: sokal@nyu.edu

Reply to comments on Brown et al. (2013)

We are gratified that our article (Brown, Sokal, & Friedman, 2013) has elicited such interest, and we are happy to reply briefly to each of the five comments.

Nickerson (this issue) concurs with our conclusion (Brown, Sokal, & Friedman, this issue) that there is no empirical evidence for any critical positivity ratio. She also makes an important contribution to this discussion by stressing the distinction between within-person across-time theories and within-time across-persons theories. Both types of theories are valuable in psychology, but they are conceptually distinct and by no means equivalent. We feel that this distinction deserves to be more widely discussed in the literature on research methods.

Guastello (this issue) argues that nonlinear-dynamics models in psychology can be tested empirically using time-series data. We concur, but we take no position on the specific tests that he advocates or on the nonlinear-dynamics models in psychology that he asserts are "supported by real data."

Musau (this issue) contends that we went "too far" in describing Fredrickson and Losada's (2005) article as "based on a series of erroneous and, for the most part, completely illusory `applications' of mathematics" (Brown et al., 2013, p. 812). But he does not present any evidence that this characterization of Fredrickson and Losada (2005) is anything other than literally accurate; indeed, he explicitly *declines* to "venture into the validity of Brown et al.'s assertions," despite having impugned the accuracy of those assertions in his immediately preceding sentence! Instead, Musau merely uses a hypothetical example to argue that mathematical modeling can, under appropriate circumstances, be valid and useful in psychology—which is something we never disputed.

Along similar lines, Hämäläinen, Luoma, and Saarinen (this issue) chide us for saying that Fredrickson and Losada (2005) "contains numerous fundamental conceptual and mathematical errors" (Brown et al., 2013, p. 801).  They concede our main point, namely the complete lack of justification for the use of the Lorenz equations in modeling the time evolution of human emotions; but they also assert, incorrectly, that Fredrickson and Losada's article contains "no clear mathematical errors."  Among the purely mathematical errors clearly noted in our article are Fredrickson and Losada's assertion that the $r = 22$ data (alleged to be characteristic of "medium-performance teams") end up in a limit cycle (Brown et al., 2013, p. 808), and their implicit claims concerning the absence of chaotic attraction at large values of $r$ (Brown et al., 2013, p. 812).  But we are happy to agree with Hämäläinen et al. that the central flaws in Fredrickson and Losada (2005) and its predecessor articles are logical and conceptual, not narrowly mathematical.  And they are, as we have demonstrated, overwhelming.

Where we unequivocally take issue with Hämäläinen et al. (this issue) is with their assertion that physicists derive their models "by fitting equations to empirical data."  In fact, the vast majority of models in contemporary physics, including virtually all those in which the mechanical functioning of a system is described by differential equations, are derivable from first principles by applying well-verified theories to a specific problem and then making plausible simplifications or approximations.  Indeed, Lorenz's (1963) article—from which Fredrickson and Losada (2005) purported to take their inspiration—is an excellent example of this process: Lorenz started from the standard equations of fluid dynamics for convective motion, and then truncated those equations by retaining only three Fourier modes (the ones that were most important to the solutions under study).  Empirical observations may serve to validate a theoretical model, and discrepancies between predicted and observed results can lead to the modification of theories or the generation of entirely new ones, but it is misguided to think that contemporary physicists' models are derived solely, or even principally, "by fitting equations to empirical data."

In the remainder of their comment, Hämäläinen et al. (this issue) venture into even more hazardous epistemological territory. They acknowledge that "in the behavioral sciences ... very few problems can be described by differential equations so that these would be fitted to empirical data successfully," but they then draw the odd conclusion that this lack of empirical confirmation is "a rationale to embrace mathematical modeling, not to discard it." One is left wondering under what set of circumstances these authors would see a rationale *not* to "embrace mathematical modeling." Finally, their statement that "When models are used to facilitate scientific reasoning, the most important issue is not the adequacy of fit with empirical data, but whether the model fits the purpose of the model at hand" leaves us, frankly, astonished. The link between theory and empirical test in science is often a subtle one, but the absolute need for such a link cannot simply be dismissed by vague claims about whether a model fits its own purpose.

Lefebvre and Schwartz (this issue) start from the empirical claim that "for flourishing subjects the ratio of positive to negative emotion is greater than 3, and for languishing subjects this ratio is less than 3." The trouble is that this assertion is entirely incorrect. Since the empirical evidence concerning the alleged critical positivity ratio of 3 is examined at length in our latest article (Brown, Sokal, & Friedman, this issue) and also briefly in Nickerson's comment (this issue), we need not repeat the details here. Suffice it to say that there is *no* evidence whatsoever for the existence of *any* critical positivity ratio (whether 3 or anything else); indeed, there is significant evidence against. So, far from this being "a verified empirical ratio in search of a theory," what Lefebvre and Schwartz present in the remainder of their comment is, rather, a theory aimed at explaining a nonexistent phenomenon. Moreover, their theory fails even at this latter task, because it does not predict any discontinuous phase transition: rather, under the (arbitrary) assumption that $P_b = 1/2$, the ratio $P/(1−P)$ passes *smoothly* through 3 as $P_a$ passes through the (arbitrary) dividing line 1/2.

So what are the implications for psychology of the demise of the critical positivity ratio?

Almost 50 years have passed since the appearance of Meehl's (1967) classic article contrasting the methods of physics and psychology. One of the issues raised by Meehl is that "there exists among psychologists ... a fairly widespread tendency to report experimental findings with a liberal use of *ad hoc* explanations for those that didn't 'pan out'" (p. 114); we are not convinced that this tendency is any less widespread today. Of course, psychologists can, with considerable justification, retort that theirs is the most difficult science of all, with a subject matter vastly more complex than that of any other field; and this complexity accounts, in an entirely honorable way, for the almost complete lack of what would in other sciences be recognized as theory— namely, that which "allows you [to] infer what would happen to things in certain situations without creating the situations" (Borsboom, 2013, para. 4). However, psychologists need to accept that they cannot have their "hard science" cake and eat it too. The mathematical models used in the physical sciences, many of which are based on differential equations, require extensive replication and confirmation—often to extremely high precision— before they are accepted; and they are put to a severe test every time any of us turns on a computer or boards an airplane.

It is, of course, possible that psychologists might one day develop quantitative models of some aspects of human behavior with sufficiently high predictive value that one could make decisions in therapeutic, public-policy, or other applied contexts with almost complete confidence in their outcomes. However, given the present limitations of our ability to predict how people are likely to behave in the next second—let alone the next hour, week, or year—it seems to us that, for the foreseeable future, any claims to have identified such a quantitative model of the psychology of the individual will merely turn out to be further examples of what could be termed "romantic scientism": namely, unfulfillable dreams for a simple "scientific" explanation of complex phenomena, combined with an inadequate appreciation of the degree of empirical confirmation that is a requisite of genuine science. The saga of the critical positivity ratio—and its largely uncritical acceptance within the positive psychology community— exemplifies both aspects. The fact that the three-dimensional plot of the Lorenz attractor bears a certain degree of visual resemblance to a butterfly when inspected from particular angles is without doubt aesthetically pleasing, but this observation does not somehow romantically imbue the series of equations that generate the attractor with the power to predict human emotions.

Seligman and Csikszentmihalyi (2000), in their founding manifesto of positive psychology, criticized its precursor, humanistic psychology, for its lack of scientific rigor and corresponding unbridled romanticism, and promised a more scientifically responsible approach. They stated:

> Unfortunately, humanistic psychology did not attract much of a cumulative empirical base, and it spawned myriad therapeutic self–help movements. … However, one legacy of the humanism of the 1960s is prominently displayed in any large bookstore: The "psychology" section contains at least 10 shelves on crystal healing, aromatherapy, and reaching the inner child for every shelf of books that tries to uphold some scholarly standard. (p. 7)

Yet Fredrickson and Losada's (2005) article, which has become one of the pillars of positive psychology and has both been widely cited in the scientific literature and influential in many applied realms, has betrayed that promise by replicating—albeit in a different style—many of the same problems that were criticized by Seligman and Csikszentmihalyi. Even the bookstores are replete with literally dozens of popular treatments of positive psychology that present the critical positivity ratio of 2.9013 (or 2.9, or 3) as a major scientific finding, when it is nothing of the sort. That the sin is now romantic scientism rather than pure romanticism is not, in our view, a great advance.

Fredrickson and Losada's (2005) article was the subject of over 350 scholarly citations before our critique (Brown et al., 2013) appeared, and its principal "conclusions" have been featured in many lectures and public presentations by senior members of the positive psychology research community, although its deficiencies ought to have been visible to anyone with a modest grasp of mathematics and a little curiosity. Unfortunately—because human behavior is, after all, complex and difficult to understand—we have no way of knowing whether the fact that it took so long for these deficiencies to be recognized was due to an unwarranted degree of optimism about the reliability of the peer-review process, a reluctance to make waves in the face of powerful interests, a general lack of critical thinking within positive psychology, or some other factor. We hope that our revelation of the problems with the critical positivity ratio ultimately demonstrates the success of science as a self-correcting endeavor; however, we would have greatly preferred it if our work had not been necessary in the first place.